\documentstyle[11pt,epsf]{article}

\begin{document}
\begin{center}
{\Large What really are the best 100m performances?} \\
\vspace{5mm}
J. R. Mureika \\
{\it Department of Computer Science \\
University of Southern California \\
Los Angeles, CA~~90089~~USA}
\end{center}

	The title of the ``Fastest Man in the World'' has always resided
in the domain of the 100m champion.  There is perhaps no greater test of
strength, power, and agility for a human being.  In late
July 1997, this title returned to its home in Canada as our own Donovan Bailey
crossed the finish line in a remarkable 9.84s, even after a relatively slow
reaction time of 0.174s and a small tail wind of +0.7 m/s.  His incredible
top speed of 12.1 m/s is further support to his claim on the title.

	However, this race was run with a non--zero
tail wind. That is, Bailey had an advantage not over his competitors,
but over 100m times from other races.  While the legal wind speed limit
is +2.0 m/s for the 100m and 200m sprints, one can never discount
the fact that a race run with a +1.9 m/s tail wind has an implicit advantage
over a race run with a 0.0 m/s tailwind, or even a headwind, for that matter.
Despite these rules, is it possible to compare {\it all} 100m races
on a more or less equal footing?

	The answer, to a degree, is ``yes'', and results from a little application
of the physics of fluid mechanics.  A runner moving through a wind with an
arbitrary velocity experiences either a resistive or propulsive force, as well
as a drag effect.  The former are the result of Newtonian mechanics (force
laws), and the drag depends on such factors as the runner's mass,
speed, cross--sectional area, and density of surrounding air.  

	It was determined that a sprinter loses bewtween 3--6\% of
his/her energy in overcoming the drag.  A simple formula to compensate for 
accompanying wind speeds was derived to calculate the equivalent still--air
(zero wind speed) race times; for the conservative limit of 3\%, this
is

\begin{equation}
t_0 \approx \left[ 1.03 - 0.03\times \left(1 - \frac{W \times t_W}{100}\right)^2\right]\times t_W~,
\label{wind}
\end{equation}
where $t_W$ is the recorded race time, W is the wind speed, and $t_0$ is
the equivalent still--air time.  You can do it with your own race times!

	So, how would the record books look if this formula was a standard
application to world class performances?  Adjustments of various sets of 
100m times (denoted by $t_0$ and $t_W$, respectively) are shown in the
accompanying tables.  For sheer comparison,  Ben Johnson's disqualified WR 
performances of 1987 and 1988 are included.

\section{World Rankings}

	One of the mose fascinating results can be found in 
Table~\ref{mworldwind}, the wind--corrected performances.   Donovan 
Bailey's 9.84s WR adjusts to a 9.88s equivalent in calm air. Meanwhile, 
Frank Fredricks' 9.86s clocking
(Lausanne, 03 Jul 1996) was run with a wind reading of $-0.4$ m/s, which after
correction surpasses Bailey's WR performance by 0.04s!  It is certainly
conceivable that, given the proper conditions, Fredricks could have claimed
the elusive title of ``World's Fastest Man'' with this race.  In fact, if
Fredricks had given this same performance in Atlanta ({\it i.e.} with a 
wind speed of +0.7 m/s), he would have crossed the line in roughly 9.81s!

\begin{figure}[h]
\begin{center}
\leavevmode
\epsfysize=400pt
\epsfbox[45 65 560 740] {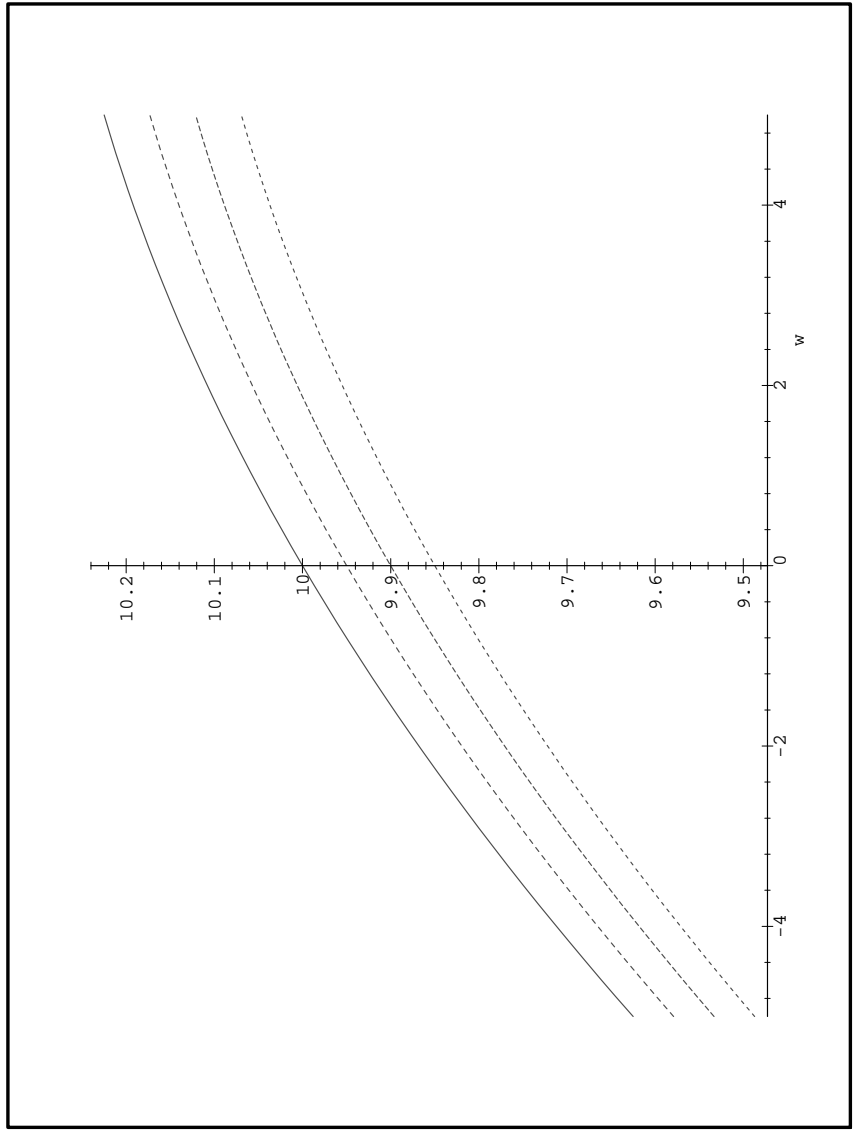}
\end{center}
\caption{Wind--correction curves for wind--assisted times of 10.00s (upper solid), 9.95 (second), 9.90 (third),
and 9.85 (bottom).  Wind speed is $w$.}
\label{windplot}
\end{figure}

	It should be noted that, due to the drag effects mentioned earlier,
races run into a head wind will have faster corresponding times than races
run with a tail wind of equivalent strength.  Figure~\ref{windplot} shows
that the ``correction curve'' is not linear, but rather a curve bending toward
the right.  Hence, a head wind will fall on the ``steeper'' portion of the
curve, while a tail wind will be on the shallower side.

	The 9.84s WR would rank 6th all--time if we included banned 
performances (Table~\ref{banned}).  After correcting for the wind conditions,
(Table~\ref{mworldwb}), this time climbs to 5th, but is surpassed by several
different performances.  Thompson's 9.69s run has a wind--corrected equivalent
of 9.93s, and has sunk to 16th.  Meanwhile, Davidson Ezinwa's 9.91s race
(into a 2.3 m/s headwind) has a wind--corrected 9.76s equivalent.  Note that
this performance is marked with a ``d'', indicating a doubtful wind 
reading\footnote{It is generally known that athletes who race at altitude 
perform better than they do closer to sea level, and it has been suggested 
that this effect may be more physiological in
nature than physical, since the corresponding change in air density does not
yield the expected change in race time.}.

	Florence Griffith--Joyner's 100m WR
performance of 10.49s at the 1988 US Olympic Trials is skeptical.  
It has been noted that,
at the time of this race (second heat of the second round), the 
wind gauge for the 100m straight read
a speed of +0.0 m/s, whereas a nearby gauge (for the jumps runway) read
speeds in excess of +4.0 m/s.  Furthermore, the wind reading for the 
third heat was +5.0 m/s.  This mysterious sudden calm was never really 
addressed; it is unreasonable to assume that the wind would
die down completely for the duration of one race.  So, assuming that the
wind speed was actually between +4.0 m/s and +5.0 m/s during Flo--Jo's
``WR'' race, she would have actually clocked a time somewhere in the 
range of 10.70s -- 10.74s for a still wind, which would be much more
consistent with her other performances (her time of 10.61s in the final was 
legal, however, with a wind speed of +1.2 m/s).

\section{Canadian Rankings}

	This analysis also shows some neat results of local interest.
For example, Bailey's 9.84s WR from Atlanta rounds to a 9.88s still--air 
equivalent. Furthermore, if the correction formula (\ref{wind}) is applied to 
Bailey's Atlanta splits\footnote{The formula has to be modified for the
distance, though; the '100' in the numerator changes to 50 and 60 to
correspond to the race distance. Although, it doesn't make much of a 
difference to leave it as 100.}, these times could be compared with indoor 
performances (where there is no wind speed) over the same distance.  In this 
case, one finds 50m and 60m times of 5.63s and 6.53s, respectively.  
The former (5.63s) is only 0.07s slower than his 50m 5.56s indoor WR (Reno, NV,
09 Feb 1996), a 
difference which could perhaps be accounted for by reaction time; {\it i.e.}
if Bailey had a reaction time of around 0.11--0.12s for his 50m WR, then
these results would be consistent. 
The latter (6.53s) is 0.02s off his 1997 indoor PB of 6.51s 
(Maebashi, Japan, 08 Feb 1997).  This would
tend to suggest that Bailey's Olympic 100m WR was consistent with his
other PB performances.

	The 1996 100m rankings can be similarly restructured.  
Table~\ref{allcan} shows the top 46 performances, accounting for wind 
assistance, and Tables~\ref{can10},\ref{canwind10} show the top 10 legal
and wind--corrected rankings.   The Canadian rankings do 
not suffer as much of a restructuring as do the World rankings.

\section{Conclusions}

	Who, then, did run the fastest 100m race ever?  Based on this model,
and discounting substance--assisted performances, doubtful wind readings,
and hand--times, Fredricks comes out the winner, and Bailey has to settle
for 2nd.  Only 3 of the top 20 performances are 
now sub--9.90, whereas before 8 out of 20 were under this mark.  The
third fastest wind--corrected time is Christie's; it would have been 
interesting had he not false--started out of the final in Atlanta.  Only
about 7 of the top 20 wind--corrected athletes will be competing this
year (who knows if Christie will be back with a vengeance?).  
We'll most likely see the most sub--9.90s races to date.  Most importantly,
though, Fredricks has the real potential to better Bailey's existing record.
Of course, Bailey will also have the same potential.  It seems quite likely
that the 100m WR drop from its 1996 mark.  Will we see a sub--9.80s race?  
If so, who will be the one to run it?  Based on their best races last year, 
Bailey could run a 9.79s with a $+$1.7m/s tailwind, while Fredricks would 
need a mere $+$1.0m/s!  With more training under their belt, who knows what to
expect. Watch for a showdown between these two at the 1997 WC!

\pagebreak
\begin{table}[t]
\begin{center}
{\begin{tabular}{|c c c c l l l|}\hline
\#& & $t_W$ & $W$ & Athlete & Location& Date \\ \hline\hline
1&&9.84 & +0.7 & Donovan Bailey       CAN & Atlanta, GA &  27 Jul 1996 \\
2&&9.85 & +1.2 & Leroy Burrell        USA & Lausanne, SWI &  06 Jul 1994 \\
3&&9.86 & $-0.4$ & Frank Fredericks     NAM & Lausanne, SWI &  03 Jul 1996 \\
4&&9.86 & +1.2 & Carl Lewis           USA & Tokyo, JAP &  25 Aug 1991 \\
5&&9.87 & +0.3 & Linford Christie     GBR & Stuttgart, GER &  15 Aug 1993   \\
6&&9.87 & +1.9 & Fredricks	NAM & Helsinki, FIN & 25 Jun 1996 \\
7&&9.88 & +1.2 & Burrell              USA & Tokyo, JAP & 25 Aug 1991  \\
8&&9.89 & +0.7 & Fredericks     NAM & Atlanta, GA & 27 Jul 1996 \\
9&&9.90 & +0.7 & Ato Boldon           TRI &Atlanta, GA&  27 Jul 1996 \\
10&&9.90 & +1.9 & Burrell	USA & New York, NY & 14 Jun 1991 \\
11&&9.91 & +1.3 & Bailey   CAN	& Montr\'{e}al, PQ & 15 Jul 1995 \\
12&&9.91 & +1.5 & Dennis Mitchell      USA &Milan, ITY &  07 Sep 1996   \\
13&&9.91 & +1.9 & Christie	GBR  & Victoria, BC  & 23 Aug 1994 \\
14&&9.92 & +0.3 & Andre Cason          USA &Stuttgart, GER&  15 Aug 1993\\
15&&9.92 & +0.8 & Boldon   TRI & Eugene, OR & 01 Jun 1996 \\
16&&9.92 & +1.1 & Lewis  USA	& Seoul, SK & 24 Sep 1988 \\
17&&9.92 & +1.1 & Mitchell USA & Atlanta, GA & 16 Jun 1996 \\
18&&9.92 & +1.2 & Christie  GBR	& Tokyo, JAP & 25 Aug 1991 \\
19&&9.93 & $-0.6$ & Mike Marsh           USA & Walnut, CA&  18 Apr 1992  \\
20&&9.93 & $-0.6$ & Boldon         TRI  & Atlanta, GA &  27 Jul 1996 \\ \hline
\end{tabular}}
\end{center}
\caption{Men's Top 20 fastest legal 100m times, ranked according to increasing
tail--wind speed (for equal time runs).}
\label{mworldlegal}
\end{table}

\pagebreak
 
\begin{table}[h]
\begin{center}
{\begin{tabular}{|c c c c c c l l|}\hline
New&Old&& $t_0$ & $t_W$ & $W$ & Athlete & Date \\ \hline\hline
1&3&   & 9.84& 9.86& $-0.4$ & Frank Fredericks  NAM  & 03 Jul 1996 \\
2&1&    & 9.88& 9.84& +0.7 & Donovan Bailey    CAN  & 27 Jul 1996 \\
3&5&    & 9.89& 9.87& +0.3 & Linford Christie  GBR  & 15 Aug 1996 \\
4&56&    & 9.89& 9.97& $-1.3$ & Leroy Burrell       USA  & 01 Aug 1992 \\
5&57& D  & 9.90& 9.97& $-1.2$ &  Ben Johnson       CAN  & 19 Aug 1987 \\
6&20&    & 9.90& 9.93& $-0.6$ &  Ato Boldon        TRI  & 27 Jul 1996 \\
8&21&    & 9.91& 9.93& $-0.4$ &     Bailey        CAN  & 03 Jul 1996 \\
9&29&    & 9.91& 9.94& $-0.5$ &     Fredericks    NAM  & 27 Jul 1996 \\
10&2&    & 9.92& 9.85& +1.2 &  Burrell     USA  & 06 Jul 1994 \\
11&30&    & 9.92& 9.94& $-0.4$ &     Boldon        TRI  & 03 Jul 1996 \\
12&46& D  & 9.93& 9.96& $-0.6$ & Davidson Ezinwa        NGR  & 18 Apr 1992 \\
13&--& w  & 9.93& 9.69& +5.7 & Obadele Thompson  BAR  & 13 Apr 1996 \\
14&4&    & 9.93& 9.86& +1.2 & Carl Lewis        USA  & 25 Aug 1991 \\
15&8&    & 9.93& 9.89& +0.7 &     Fredericks    NAM  & 27 Jul 1996 \\
16&37&    & 9.94& 9.95& $-0.3$ &     Boldon        TRI  & 27 Jul 1996 \\
17&14&    & 9.94& 9.92& +0.3 & Andre Cason       USA  & 15 Aug 1993 \\
18&9&    & 9.94& 9.90& +0.7 & Boldon        TRI  & 27 Jul 1996\\
19&7&    & 9.95& 9.88& +1.2 &  Burrell       USA  & 25 Aug 1991 \\
20&47&    & 9.96& 9.96& $-0.1$ &     Cason         USA  & 14 Aug 1993 \\ \hline
\end{tabular}}
\end{center}
\caption{Men's Top 20 fastest wind--corrected 100m times, including
wind--aided performances (w) and legal times from athletes caught for
doping during their career (D).}
\label{mworldwind}
\end{table}

\begin{table}
\begin{center}
{\begin{tabular}{|c c r r l l l|}\hline
Rank& &$t_W$ & $W$ & Athlete &Location& Date \\ \hline
1&w&9.69& +5.7 & Obadele Thompson     BAR &El Paso,TX&  13 Apr 1996 \\
2&w&9.78& +5.2 & Carl Lewis           USA &Indianapolis, IN&  16 Jul 1988 \\
3&B&9.79& +1.1 & Ben Johnson          CAN &Seoul, SK&  24 Sep 1988 \\
4&w&9.79& +4.5 & Andre Cason          USA &Eugene, OR&  16 Jun 1993 \\
5&B&9.83& +1.0 & Johnson   CAN  &Rome, ITY& 30 Aug 1987 \\
6&w&9.85& +4.8 & Dennis Mitchell      USA &Eugene, OR&  17 Jun 1993 \\
7&w&9.87& +11.2 & William Snoddy      USA &Dallas, TX&  01 Apr 1978 \\
8&w&9.87& +4.9 & Calvin Smith         USA &Indianapolis, IN&  16 Jul 1988 \\
9&w&9.88& +2.3 & James Sanford        USA & Westwood, CA& 03 May 1980 \\
10&w&9.88& +4.0 & Bailey   CAN   &Duisburg, GER& 12 Jun 1996 \\
11&w&9.88& +5.2 &Albert Robinson USA  & Indianapolis, IN& 16 Jul 1988 \\ 
12&w&9.88& +5.3 &Maurice Greene  USA  & Austin, TX &08 Apr 1995 \\ 
13&w&9.89& +2.9 &Mike Marsh USA &Walnut, CA& 14 Apr 1995 \\
14&w&9.89& +4.1 &Frank Fredericks     NAM &Tokyo, JAP &24 Aug 1991 \\  
15&w&9.89& +4.2 &Raymond Stewart   JAM &Indianapolis, IN& 09 Aug 1987 \\ 
16&w&9.90& +3.7 &Johnson          CAN & Ottawa, ON& 06 Aug 1988 \\
17&w&9.90& +5.2 &Joe DeLoach          USA &Indianapolis, IN&16 Jul 1988 \\
18&d&9.91& $-2.3$ &Davidson Ezinwa      NGR  &Azusa, CA&11 Apr 1992\\
19&f&9.91& +1.2 &Mitchell      USA  & Tokyo, JAP &25 Aug 1991 \\ 
20&w&9.91& +4.2 &Mark Witherspoon  USA  & Indianapolis, IN& 09 Aug 1987 \\ \hline
\end{tabular}}
\end{center}
\caption{Men's Top 20 fastest illegal 100m times, including
wind--aided performances (w), doubtful wind readings (d), false starts (f).
Ben Johnson's disqualified WR performances of 1987 and 1988 are shown for 
comparison.}
\label{banned}
\end{table}

\begin{table}
\begin{center}
{\begin{tabular}{|c c r r r l l|}\hline
Rank& &$t_0$ &$t_W$ & $W$ & Athlete &  Date \\ \hline
1& d  & 9.76& 9.91& $-2.3$ & Davidson Ezinwa   NGR        & 11 Apr 1992 \\
2& h  & 9.80& 9.70& +1.9 & Donovan Powell    JAM        & 19 May 1995 \\
3&    & 9.84& 9.86& $-0.4$ & Frank Fredericks  NAM        & 03 Jul 1996 \\
4& B  & 9.85& 9.79& +1.1 & Ben Johnson       CAN        & 24 Sep 1988 \\
5&    & 9.88& 9.84& +0.7 & Donovan Bailey    CAN        & 27 Jul 1996 \\
6& B  & 9.89& 9.83& +1.0 &  Ben Johnson   CAN          & 30 Aug 1987 \\
7&    & 9.89& 9.87& +0.3 & Linford Christie  GBR        & 15 Aug 1993 \\
8&    & 9.89& 9.97& $-1.3$ & Leroy Burrell       USA        & 01 Aug 1992 \\
9&    & 9.90& 9.93& $-0.6$ & Ato Boldon        TRI        & 27 Jul 1996 \\
10&    & 9.90& 9.93& $-0.6$ &     Boldon        TRI        & 27 Jul 1996 \\
11& D  & 9.90& 9.97& $-1.2$ &     Johnson       CAN        & 19 Aug 1987 \\
12&    & 9.91& 9.93& $-0.4$ &     Bailey        CAN        & 03 Jul 1996\\
13&    & 9.91& 9.94& $-0.5$ &     Fredericks    NAM        & 27 Jul 1996 \\
14&    & 9.92& 9.85& +1.2 & Burrell     USA        & 06 Jul 1994 \\
15&    & 9.92& 9.94& $-0.4$ &     Boldon        TRI        & 03 Jul 1996 \\
16& w  & 9.93& 9.69& +5.7 & Obadele Thompson  BAR        & 13 Apr 1996 \\
17& D  & 9.93& 9.96& $-0.6$ &     Ezinwa        NGR        & 18 Apr 1992 \\
18&    & 9.93& 9.86& +1.2 & Carl Lewis        USA        & 25 Aug 1991 \\
19&    & 9.93& 9.89& +0.7 &     Fredericks    NAM        & 27 Jul 1996 \\
20&    & 9.94& 9.95& $-0.3$ &     Boldon        TRI        & 27 Jul 1996 \\ \hline
\end{tabular}}
\end{center}
\caption{Men's Top 20 fastest wind--corrected 100m times, including
wind--aided performances (w), doubtful wind readings (d), legal performances
of athletes caught for doping during their career (D), and disqualified 
times from athletes caught for doping during their career (B).}
\label{mworldwb}
\end{table}
\begin{table}
\begin{center}
{\begin{tabular}{|c c c c c l c|}\hline
\#& & $t_0$ & $t_w$ & $v_w$ & Athlete & Date \\ \hline\hline 
1& w  & 10.67& 10.54& +2.1 &Florence Gr
iffith--Joyner  USA        & 25 Sep 1988 \\
2&    & 10.69& 10.62& +1.0 &       Griffith--Joyner     USA        & 24 Sep 1988 \\
3&    & 10.69& 10.61& +1.2 &  Griffith--Joyner  USA        & 17 Jul 1988 \\
4& d  & 10.70& 10.49& +4.0 &  Griffith--Joyner  USA        & 16 Jul 1988 \\
5&    & 10.75& 10.82& $-1.0$ & Gail Devers          USA        & 01 Aug 1992 \\
6&    & 10.76& 10.83& $-1.0$ & Juliet Cuthbert      JAM        & 01 Aug 1992 \\
7&    & 10.77& 10.84& $-1.0$ & Irina Privalova      RUS        & 01 Aug 1992 \\
8&    & 10.79& 10.86& $-1.0$ & Gwen Torrence       USA        & 01 Aug 1992\\
9&    & 10.80& 10.87& $-1.0$ & Merlene Ottey          JAM        & 07 Aug 1991\\
10&    & 10.80& 10.82& $-0.3$ &       Ottey          JAM        & 16 Aug 1993 \\
11&    & 10.81& 10.70& +1.6 &       Griffith--Joyner     USA        & 17 Jul 1988\\
12&    & 10.81& 10.84& $-0.5$ &       Devers         USA        & 23 Aug 1996 \\
13&    & 10.81& 10.78& +0.4 &       Ottey          JAM        & 03 Sep 1994 \\
14&    & 10.83& 10.74& +1.3 &  Ottey        JAM        & 07 Sep 1996\\
15&    & 10.83& 10.77& +0.9 &  Privalova      RUS        & 06 Jul 1994\\
16& A  & 10.84& 10.79& +0.6 & Evelyn Ashford        USA        & 03 Jul 1983 \\
17&    & 10.84& 10.84& +0.0&      Ottey          JAM        & 10 Jul 1991 \\
18& A & 10.85& 10.78& +1.0 & Dawn Sowell          USA        & 03 Jun 1989\\
19&    & 10.85& 10.82& +0.4 &  Torrence        USA        & 03 Sep 1994\\
20&    & 10.85& 10.85& &      Torrence       USA        & 18 May 1996\\ \hline
\end{tabular}}
\end{center}
\caption{Women's Top 20 fastest wind--corrected 100m times, including
only wind--aided and altitude (A) performances.  Griffith--Joyner's
10.54s clocking has been assigned a wind reading of +2.1 m/s, since
no accurate wind reading was found.}
\end{table}

\begin{table}
\begin{center}

{\begin{tabular}{|c c r r l l l|}\hline
Rank& &$t_W$  & $W$ & Athlete & Location & Date \\ \hline
1&&   9.84 & +0.7 & Donovan Bailey  & Atlanta, GA    & 27 Jul 1996 \\
2&&  10.03 & +0.7 & Bruny Surin     & Paris, FRA     & 29 Jun 1996 \\
3&&  10.16 & +1.9 & Glenroy Gilbert & Duisberg, GER  & 12 Jun 1996 \\
4&A&  10.18& +1.5 & Robert Esmie    & Colorado Sp.,CO& 08 Jun 1996 \\
5&&  10.19 & +0.8 & Carlton Chambers& Eugene, OR     & 29 May 1996 \\
6&&  10.30 & +0.0 & Andy Breton     & Kalamazoo, MI  & 04 May 1996 \\
7&&  10.37 & +1.1 & Peter Ogilvie   & Austin, TX     & 04 May 1996 \\
8&A&  10.40& +0.5 & Anthony Wilson  & Flagstaff, AZ  & 11 May 1996 \\
9&&  10.41 & +1.0 & Dave Tomlin     & Montreal, PQ   & 21 Jun 1996 \\
10&& 10.43 & +1.8 & Okiki Akinremi  & Eich/Abbost, BC& 02 Jun 1996 \\
&w&    9.88& +4.0 & Donovan Bailey  & Duisburg, GER  & 12 Jun 1996\\
&w&    9.97& +2.1 & Bailey          & Atlanta, GA    & 18 May 1996\\
&w&   10.08& +3.4 & Robert Esmie    & Kitchener, ON  & 09 Aug 1996\\
&w&   10.39& +4.4 & Esmie           & Kitchener, ON  & 09 Aug 1996\\
&w&   10.13& +4.0 & Glenroy Gilbert & Duisburg, GER  & 12 Jun 1996\\
&w&   10.18& +3.1 & Carlton Chambers& College Park, MD& 19 Apr 1996\\
&w&   10.27& +3.4 & Dave Tomlin     & Kitchener, ON  & 09 Aug 1996\\ \hline
\end{tabular}}
\end{center}
\caption{Canadian Men's Top 10 fastest official 100m times,including 
altitude performances (A); wind--aided times are unranked.}
\label{can10}
\end{table}
\begin{table}
\begin{center}
{\begin{tabular}{|c c c r r r l l l|}\hline
New&Old& & $t_0$ &$t_W$ & $W$ & Athlete & Location & Date \\ \hline
1& 1&& 9.88& 9.84& 0.7&   Donovan Bailey   & Atlanta, GA     & 27 Jul 1996 \\
2& 2&& 10.03& 10.05& $-0.4$&   Surin            & Lausanne, SWI   &03 Jul 1996 \\
3& 3&&10.18& 10.18& 0.0&   Gilbert          & Montr\'{e}al, PQ   & 21 Jun 1996 \\
4& 5&&10.22& 10.22& 0.0&   Chambers         & Montr\'{e}al, PQ   & 21 Jun 1996 \\
5& 4&w& 10.26& 10.08& 3.4&   Robert Esmie     & Kitchener, ON  & 09 Aug 1996 \\
6& 6&& 10.30& 10.30& 0.0&   Andy Breton      & Kalamazoo, MI   & 04 May 1996 \\
7& 8&A& 10.44& 10.40& 0.5&   Anthony Wilson   & Flagstaff, AZ& 11 May 1996 \\
8& 7&& 10.44& 10.37& 1.1&   Peter Ogilvie    & Austin, TX      & 04 May 1996 \\
9& 11&& 10.44& 10.44& 0.0&   Abass Tanko      & Sherbrooke, PQ  & 08 Jun 1996 \\
10& 9&w& 10.45& 10.27& 3.4&   Dave Tomlin      & Kitchener, ON   & 09 Aug 1996 \\ \hline
\end{tabular}}
\end{center}
\caption{Canadian Men's Top 10 fastest wind--corrected 100m times,including 
altitude performances (A).}
\label{canwind10}
\end{table}
\begin{table}
\begin{center}
{\begin{tabular}{|c c r r r l l r|}\hline
Rank& & $t_0$ &$t_W$ & $W$ & Athlete & Location & Date \\ \hline
1& & 9.88& 9.84& 0.7&   Donovan Bailey   & Atlanta, GA     & 27 Jul 1996 \\
2&& 9.91& 9.93& $-0.4$&   Bailey           & Lausanne, SWI   & 3 Jul 1996 \\
3&& 9.97& 10.00& $-0.5$&   Bailey           & Atlanta, GA     & 27 Jul 1996 \\
4&& 9.98& 9.98& +0.0&   Bailey           & Montr\'{e}al, PQ    & 21 Jun 1996 \\
5&& 10.03& 10.05& $-0.4$&   Surin            & Lausanne, SWI   & 3 Jul 1996 \\
6&& 10.04& 9.95& 1.5&   Bailey           & Milan, ITA      & 7 Sep 1996 \\
7&& 10.04& 10.04& 0.0&   Surin            & Montr\'{e}al, PQ   & 21 Jun 1996 \\
8&& 10.06& 10.13& $-1.2$&   Surin            & Atlanta, GA    & 26 Jul 1996 \\
9&w& 10.07& 9.88& 4.0&   Donovan Bailey   & Duisburg, GER   & 12 Jun 1996 \\
10&& 10.08& 10.03& 0.7&   Bruny Surin      & Paris, FRA     & 29 Jun 1996 \\
11&w& 10.09& 9.97& 2.1&   Bailey           & Atlanta, GA    & 18 May 1996 \\
12&& 10.18& 10.18& 0.0&   Gilbert          & Montr\'{e}al, PQ   & 21 Jun 1996 \\
13&& 10.20& 10.11& 1.5&   Surin            & Milan, ITA     & 7 Sep 1996 \\
14&& 10.22& 10.22& 0.0&   Chambers         & Montr\'{e}al, PQ   & 21 Jun 1996 \\
15&& 10.24& 10.19& 0.8&   Carlton Chambers & Eugene, OR      & 29 May 1996 \\
16&w& 10.26& 10.08& 3.4&   Robert Esmie     & Kitchener, ON  & 9 Aug 1996 \\
17&& 10.26& 10.22& 0.6&   Gilbert          & Hechtel, BEL   & 6 Jul 1996 \\
18&A& 10.27& 10.18& 1.5&   Robert Esmie     & Colorado Sp.,CO& 8 Jun 1996 \\
19&& 10.27& 10.16& 1.9&   Glenroy Gilbert  & Duisberg, GER   & 12 Jun 1996 \\
20&& 10.28& 10.28& 0.0&   Esmie            & Montr\'{e}al, PQ   & 21 Jun 1996 \\
21&& 10.28& 10.28& 0.0&   Esmie            & Montr\'{e}al, PQ   & 21 Jun 1996 \\
22&& 10.30& 10.30& 0.0&   Andy Breton      & Kalamazoo, MI   & 4 May 1996 \\
23&w& 10.33& 10.13& 4.0&   Glenroy Gilbert  & Duisburg, GER   & 12 Jun 1996 \\
24&& 10.34& 10.23& 1.8&   Gilbert          & Stockholm, SW   & 8 Jul 1996 \\
25&w& 10.35& 10.18& 3.1&   Carlton Chambers & College Park,MD & 19 Apr 1996 \\
26&& 10.40& 10.40& 0.0&   Esmie            & Montr\'{e}al, PQ  & 21 Jun 1996 \\
27&& 10.42& 10.35& 1.0&   Chambers         & Montr\'{e}al, PQ & 21 Jun 1996 \\
28&A& 10.44& 10.40& 0.5&   Anthony Wilson   & Flagstaff, AZ& 11 May 1996 \\
29&& 10.44& 10.37& 1.1&   Peter Ogilvie    & Austin, TX      & 4 May 1996 \\
30&& 10.44& 10.44& 0.0&   Abass Tanko      & Sherbrooke, PQ  & 8 Jun 1996 \\
\end{tabular}}
\end{center}
\end{table}
\pagebreak
\begin{table}
\begin{center}
{\begin{tabular}{|r c r r r l l r|}
31&w& 10.45& 10.27& 3.4&   Dave Tomlin      & Kitchener, ON   & 9 Aug 1996 \\
32&& 10.46& 10.46& $-0.1$&   Esmie            & Kerkrade, HOL   & 17 May 1996 \\
33&& 10.48& 10.41& 1.0&   Dave Tomlin      & Montr\'{e}al, PQ & 21 Jun 1996 \\
34&& 10.54& 10.43& 1.8&   Okiki Akinremi   & Eich/Abbost, BC & 2 Jun 1996 \\
35&& 10.55& 10.52& 0.4&   Esmie            & Stockholm, SW   & 8 Jul 1996 \\
36&& 10.57& 10.57& 0.0&   Bradley McCuaig  & Montr\'{e}al, PQ  & 21 Jul 1996 \\
37&& 10.58& 10.55& 0.4&   Trevino Betty    & Montr\'{e}al, PQ & 21 Jun 1996 \\
38&& 10.59& 10.48& 1.8&   Chambers         & Stockholm, SW   & 8 Jul 1996 \\
39&& 10.59& 10.59& 0.0&   Sheridan Baptiste& Baton Rouge, LA & 11 May 1996 \\
40&& 10.60& 10.49& 1.8&   Ricardo Greenidge& Rich/Abbost, BC & 2 Jun 1996 \\
41&& 10.61& 10.61& $-0.1$&   Kofi Yevakpor    & Sherbrooke, PQ  & 15 Jul 1996 \\
42&w& 10.61& 10.39& 4.4&   Esmie            & Kitchener, ON   & 9 Aug 1996 \\
43&& 10.63& 10.60& 0.4&   Charles Allen    & Kitchener, ON   & 7 Jun 1996 \\
44&& 10.67& 10.64& 0.4&   Bryce Coad       & Kitchener, ON   & 7 Jun 1996 \\
45&& 10.70& 10.63& 1.0&   Troy Dos Santos  & Montr\'{e}al, PQ    & 21 Jun 1996 \\
\hline
\end{tabular}}
\end{center}
\caption{Canadian Men's Top fastest wind--corrected 100m times,including altitude performances (A).} 
\label{allcan}
\end{table}

\end{document}